\magnification=\magstephalf
\baselineskip=12pt
\vsize=22.0truecm
\hsize=15.5truecm 
\nopagenumbers
\parskip=0.2truecm
\def\vs{\vskip 0.2in}
\def\ts{\vskip 0.05in}

\def\n{\noindent}

\def\deg{^\circ}
\font\bigbold=cmbx12

\centerline{{\bigbold Spatially Filtering the Binary Confusion Noise}}

\centerline{{\bigbold for Space Gravitational Wave Detectors}}
\vs
\centerline {Ronald W. Hellings}
\centerline {Montana State University}
\centerline {Bozeman MT 59715 }

\vs
\n ABSTRACT: For the last fifteen years, the limiting noise source at the low frequency end of the sensitivity window for space gravitational wave detectors has been expected to be the confusion background of overlapping galactic binary stars.  Here, we present results of a study that investigates the correlation between binary star signals and conclude that there is a spatial filter in the position-dependent Doppler shift of each binary that sharply reduces the contribution of the galactic binary confusion to the noise in the detector when a monochromatic source is being detected.  The sensitivity is thus determined by the instrument alone, and the confusion noise may effectively be ignored.

\vfill
\eject

It has been nearly 15 years since the first paper [1] was published recognizing the problem that thousands of compact binaries would present in the detection of low frequency (LF) gravitational waves.  Since that time, all discussions of the sensitivity of the detectors to gravitational waves have had to take these binaries into account, concluding that at periods longer than about 300 seconds the detectors will likely be limited not by the instruments themselves, but by the inability of the detectors to separate a weak gravitational wave of interest from the background of the gravitational waves from these other, less interesting sources.  In this paper we show, for the case where the source of interest is itself a particular monochromatic binary, that there is an inherent spatial filter due to the phase modulation produced by the motion of the detector around the sun that sharply reduces the confusion noise, typically to a level below the instrument noise of the detector.  Therefore, for monochromatic sources like these, the confusion noise may simply be ignored.

We will consider the two different configurations of space gravitational detectors that have been proposed.  These are the ecliptic plane configuration (like the proposed OMEGA [2] detector in which independent probes orbit the earth in the ecliptic plane) and the precessing plane case (like the proposed LISA [3] mission where the heliocentric orbits for the probes are chosen so that the plane of the detector is inclined $60\deg$ to the ecliptic and precesses around the ecliptic pole once per year).  A binary star of total mass $M$ and reduced mass $\mu$, with circular orbit at an angular frequency $\omega$, lying in a direction given by angular coordinates $\Theta$ and $\Phi$ (measured in the plane of the detector) and with the orbit plane inclined by $i$ to the line-of-sight, will produce a signal in the detector given by
$$ \eqalign{
h(t)=&-{\sqrt{3}G^{5/3}\over c^4}\mu M^{2/3}\omega^{2/3}\cr
&\times\Bigg\{
(1+\cos^2i) \Big[{1\over2}(1+\cos^2\Theta)\cos2\Phi\cos2\Psi
 -\cos\Theta\sin2\Phi\sin2\Psi\Big]\cos(2\phi_r)\cr
 &\quad+2\cos i\Big[{1\over2}(1+\cos^2\Theta)\cos2\Phi\sin2\Psi
 +\cos\Theta\sin2\Phi\cos2\Psi\Big]\sin(2\phi_r)\Bigg\} ,
}   \eqno{(1)} $$
where $\phi_r$ is orbital phase at reception, given by
$$ \phi_r=\omega t-{\omega R\over c}\cos\Omega t , \eqno{(2)}$$
with $R$ as the radius of the detector's orbit around the sun and $\Omega$ as the detector's mean motion about the sun in radians per second.  In Eq.\ (1), $\Psi$ is the angle, measured in the plane perpendicular to the line-of-sight to the binary, between the detector plane and the major axis of the apparent ellipse of the binary's circular orbit in the sky.

Both space gravitational wave detectors that have been proposed (OMEGA and LISA) use a detector configuration in which laser signals are passed along three nearly-equal arms, the three arms forming the sides of an equilateral triangle (see Figure 1). When the signals from two adjacent arms are combined, a Michelson interferometer is formed.  Three interferometers may thus be formed, one at each vertex where the adjacent arms come together, but only two of these provide independent information.  In both configurations, the probes rotate in the plane of the detector, at a rate of once per year for the precessing plane case and at the orbital frequency of the probes for the ecliptic plane case.  Data from the two independent interferometers that are formed in each configuration may be combined in a time-dependent linear combination to simulate two non-rotating detectors, each of which is sensitive to a particular gravitational wave polarization.  To simplify the correlation of signals from different directions, we will eliminate the signal modulation produced by the in-plane rotation of the detector and will consider a single non-rotating interferometer responding to a single polarization.  The information content with this simplification is unchanged from that in the rotating case. 

For the ecliptic plane configuration, the relationship between the detector coordinates of the binary, $\Theta$ and $\Phi$, and its ecliptic coordinates, $\theta$ and $\phi$, is simply
$$ \Theta=\theta \quad\quad {\rm and} \quad\quad \Phi=\phi+\alpha, \eqno{(3)}$$
where $\alpha$ is the angle between the detector centerline and the zero of ecliptic longitude at the vernal equinox.  For the precessing plane case the situation is more complicated.  The relationship between $\{\Theta,\Phi\}$ and $\{\theta,\phi\}$ is
$$ \cos\Theta={1\over2}\cos\theta-{\sqrt{3}\over2}\sin\theta\cos\beta 
\eqno{(4a)} $$
and  
$$ \Phi=\tan^{-1}\left[{\sqrt{3}\cos\theta+\sin\theta\cos\beta\over
2\sin\theta\sin\beta}\right]  , \eqno{(4b)}$$
where $\beta$ is the time-dependent longitude of the center of the detector in the ecliptic plane, $\beta=\Omega t$. In addition, the relationship between $\Psi$ in the detector plane and the polarization angle $\psi$ in the ecliptic plane is $\Psi=\psi$ in the ecliptic plane case and
$$ \tan\Psi=
{-\sqrt{3}\sin\beta\cos\psi+(\sqrt{3}\cos\theta\cos\beta+\sin\theta)\sin\psi
\over
\sqrt{3}\sin\beta\sin\psi+(\sqrt{3}\cos\theta\cos\beta+\sin\theta)\cos\psi} 
\eqno{(4c)}  $$
for the precessing plane case.

The key to the spatial filter we are analyzing is Eq.\ (2), the position-dependent phase modulation of each binary's signal by an amount $R\omega/c$ at frequency $\Omega$.  Although the gravitational wave from a binary star is monochromatic with a fixed polarization, the motion of the detector and, in the precessing plan case, the rotation of the detector plane produce a complicated signal in the detector.  A simple Fourier power spectrum, therefore, will not detect individual sources.  To detect a binary, one must disentangle the complicated spectrum by estimating all of the relevant parameters.  This parameter estimation process is equivalent to a cross-correlation of the signal with a set of parameter-dependent templates, looking for the template that produces significant correlation.  In what follows, we present the results of numerical simulations of the correlation of a signal from a particular direction with a background of signals from many directions.

First, to demonstrate the effect, we consider a single source in the ecliptic at  $\phi=0$, with inclination $i=90\deg$ and polarization angle $\psi=0$, and correlate this signal, first with itself and then with other sources with identical parameters but at varying ecliptic longitudes.  We use Eqs. (1), (2), and (3) or (4) to generate the signals.  The computer code used to generate these signals was derived from that used in Moore and Hellings [4].  To illustrate the dependence of the spatial filtering on frequency, we calculate the correlations for sources and backgrounds at frequencies from $10^{-4}$Hz to $10^{-2}$Hz. Results of these simulations are shown in Figs. 2a and 2b.

In Figs. 2, the comparison sources all had the same inclination and polarization as the reference source at $\phi=0$.  This should produce the greatest correlated signal for comparison sources that lie in nearly the same direction as the reference source.  However, as sources are found further away on the celestial sphere, different inclinations and polarizations may actually correlate more strongly.  Thus, we would expect a set of comparison sources with random directions, inclinations, polarizations, and phases to have correlations that lie {\it inside} the envelope of the Fig.\ 2 curves when the sources are close to the reference source, but that often lie {\it outside} it when the sources are further away.  Figs. 3a and 3b illustrate this.  For these figures, a single $f=3\times10^{-3}$ Hz reference source was chosen in the ecliptic plane at $\phi=0$, and 250 comparison sources, each at the same frequency and intrinsic amplitude but with other parameters randomly chosen to provide even distribution over the sky, were correlated with the reference source.  The simple sum of all the correlations in each case is given in each figure.  Thus, for example, the correlation of the sum of all 250 sources with the single source in the ecliptic plane case (Fig.\ 2a) is only 2.23 times the amplitude of the single source itself.  Since the amplitude of the 250 combined sources is $\sim\sqrt{250}\approx 16$ times the amplitude of a single source, one would expect to be able to see a single source whose amplitude is $2.23/\sqrt{250}=0.141$ lower than the confusion noise.  Although these numbers are exact only for the particular example displayed in the figures, they are typical of the $\sim$75 random cases that were run.

WE have found that the two configurations (ecliptic-plane and precessing-plane) have nearly the same average correlations over the whole sky.  The precessing-plane case is fairly uniform over the sky, while the ecliptic plane case is somewhat worse near the ecliptic plane (the 2.23 in Fig.\ 3a compared with 1.30 in Fig.\ 3b) and somewhat better for a source near the pole (See Fig.\ 4). 

Finally, we have simulated the actual data analysis process. We have generated a signal made up of the combined signals from all the sources used to produce Fig.\ 3 and have added to it a single source, stronger than the individual sources that made up the background but weaker than the sum of the 250 background sources.  That total signal was then correlated with a template representing the source and the correlation with the correct template was compared with that from other templates.  We have chosen the precessing plane configuration for the simulation.  Seen in the detector, the {\it rms} amplitude of the confusion background was $7.6\times10^{-22}$, while that of the source was $5.5\times10^{-22}$.  The spectral content of the confusion background and of the source are shown in Fig.\ 5, where the complicated spectrum in the moving detector of what is inherently a pure monochromatic signal at $3\times10^{-3}$Hz is evident.

The data recovery simulation then proceeded as follows.  Three sources were taken, all in the ecliptic plane, one at $\phi=0$, one at $\phi=60\deg$, and one at $\phi=90\deg$.  These were added separately to the confusion background to produce three signals. Each of the three signals were then cross-correlated with  three templates, each template appropriate to one of the sources.  Results are shown in Fig.\ 6.  In Fig.\ 6a, for example, the signal contained the source at $\phi=0$.  It was then correlated with templates for $\phi=0$, $\phi=60\deg$, and $\phi=90\deg$.  The strong correlation with the proper template compared with the correlation with the other templates is seen clearly in the figure.  Similar plots are shown for the other two sources in Figs. 5b and 5c.  We reiterate that the particular sources in each case actually have amplitudes lower than the background, but have correlated amplitudes well above it.

We conclude with a few remarks about the choices we have made for this study.  First, in generating Figs. $3-6$, we have used sources with frequency $3\times10^{-3}$Hz.  This is the presumed upper edge of the frequencies where the confusion background spectrum dominates the instrument noise spectrum.  If the presumptions about the white dwarf binary background are correct, then the spatial filter does not enhance sensitivity above this frequency.  However, other assumptions about the white dwarf binary background [5,6] would produce a confusion noise at frequencies well above this.  In this case confusion noise would dominate over much more of the sensitivity window, and the spatial filter becomes even more important (and is more effective). Second, at frequencies below $3\times10^{-3}$Hz, the factor by which the spatial filter suppresses noise is reduced, the average factor of 10 at $3\times10^{-3}$Hz falling to an average factor of 5 at $1\times10^{-3}$Hz and dropping to about unity ({\it i.e.} no spatial filtering) at $1\times10^{-4}$Hz. Thus, comparing the residual binary confusion noise to the estimated instrument noise for, say, LISA [3], we see that the confusion noise drops below the instrument noise across most of the band from $10^{-4}$Hz to $10^{-2}$Hz, the only exception being in a small range around $10^{-3}$Hz where the residual binary noise does remain slightly above the instrument noise.  Finally, our choice to use 250 sources, even at the higher frequencies where the number of sources would be much lower, was made in order to get the correct statistical results.  That is, the probability of a single strong source lying in a particular direction is small, so the amplitude times the probability for each element of a statistical ensemble is equivalent to a weak source, and the ensemble average is well-modeled by adding many weak sources, as we have done in this study.  Of course, one might be unlucky and the one or two competing noise sources might lie close in the sky to the source of interest, but an unbiased prediction of sensitivity should not give undue weight to this worst case.

We reiterate that the results we have found in this study apply to detection of a single monochromatic binary in the presence of the presumed galactic binary confusion noise.  In a future paper, we will consider the application of these ideas to the detection of a chirping and coalescing binary over the same background.

The author would like to thank Tom Moore for discussions on modifying the previous code. This work was supported by NASA grant NCC5-579.

\vs
\vs

\n {\bf References:}
\ts

\font\bo=cmbx10

\n
\hangindent=0.2in
\hangafter=1
[1] D. Hils, P.L. Bender, J.E. Faller, and R.F. Webbink, in {\it 11th Int. Conf. on General Relativity and Gravitation: Abstracts of Contributed Papers, Vol. 2} (University of Stockholm, 1986), 509.

\hangindent=0.2in
\hangafter=1
\n [2]  R.W. Hellings {\it et al.}, {\it Orbiting Medium Explorer for Gravitational Astrophysics (OMEGA)}, proposal to NASA Medium Explorer program (1998, unpublished).

\hangindent=0.2in
\hangafter=1
\n [3]  P. Bender {\it et al.}, {\it LISA Pre-Phase A Report (second edition)} (1998).

\n 
[4] T.A. Moore and R.W. Hellings, Phys. Rev. D, to be published (2001).

\n 
[5] E.D. Webbink, Ap. J. {\bo 277} 355-360 (1984).

\n [6] 
W.A. Hiscock {\it et al.}, Ap. J. Lett. {\bo 540} L5-L8 (2000).

\vfill
\eject
\n CAPTIONS:

\n Figure 1.  The geometry of the source position in the plane of the detector. The two legs forming a single Michelson interferometer are shown, the centerline of the detector forming the $x$-axis of the detector coordinate system.  The binary orbit plane depicted here is at inclination $i=0$.

\n Figure 2.  Correlation between a reference source at $\phi=0$ (with $i=\psi=0$, and $\theta=90\deg$) and other sources with varying $\phi$, but with all other parameters the same.  Four curves are shown in each figure, one for each of the chosen frequencies ($10^{-2}$Hz, $3\times10^{-3}$Hz, $10^{-3}$Hz, and $10^{-4}$Hz).  Figure 2a is for the ecliptic plane configuration and Figure 2b is for the precessing plane configuration.

\n Figure 3.  Correlation between a reference source at $\phi=0$ (with $f=3\times10^{-3}$, $i=\psi=0$, and $\theta=90\deg$) and 250 other sources with the same frequency, but with random values for all other parameters.  Figure 3a is for the ecliptic plane configuration and Figure 3b is for the precessing plane configuration. 

\n Figure 4. Correlation, for the ecliptic plane case, between a reference source at $\theta=0$ (with $f=3\times10^{-3}$and $i=\psi=0$) and 250 other sources with the same frequency, but with random values for all other parameters.

\n Figure 5. Periodograms for: (a) the background of 250 random sources with $f=3\times10^{-3}$Hz and (b) a single source at $\phi=0$, $f=3\times10^{-3}$Hz, $i=\psi=0$, and $\theta=90\deg$.  The horizontal axis gives the number of frequency bins, each of width $\Delta f=1/{\rm yr}$, offset from the fundamental $f=3\times10^{-3}$Hz.

\n Figure 6.  The results of cross-correlation of templates with a signal composed of binary confusion noise plus a single source in the ecliptic plane at: (a) $\phi=0$, (b) $\phi=60\deg$, or (c) $\phi=90\deg$.  Three templates are used for each signal, representing sources at $0$, $60\deg$, and $90\deg$, labeled T00, T60, and T90, respectively.

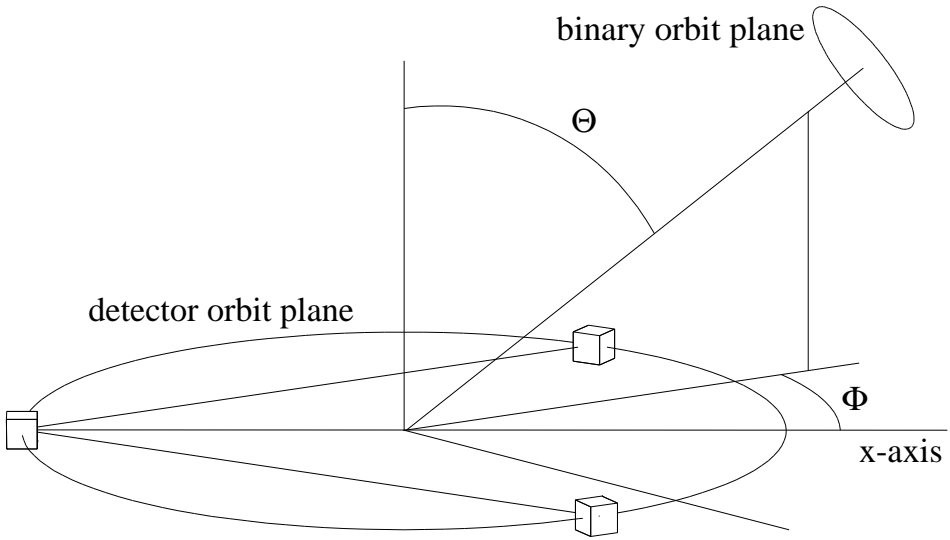
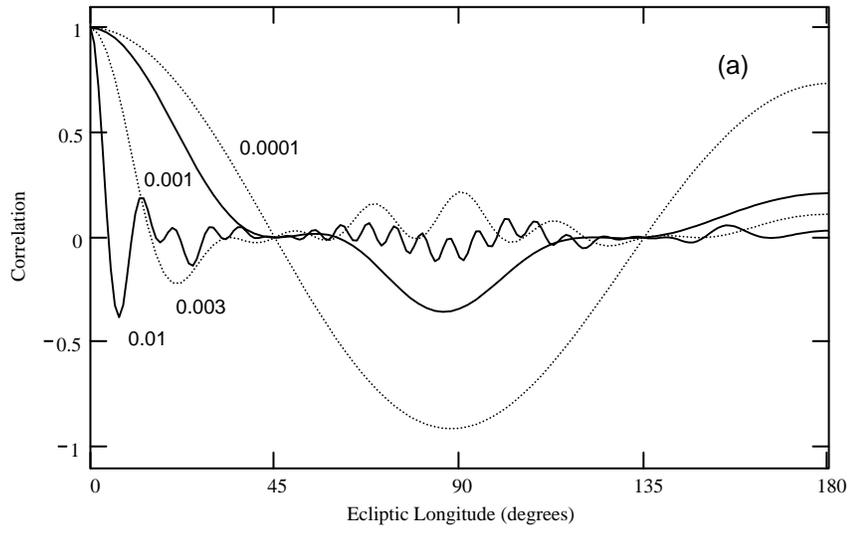
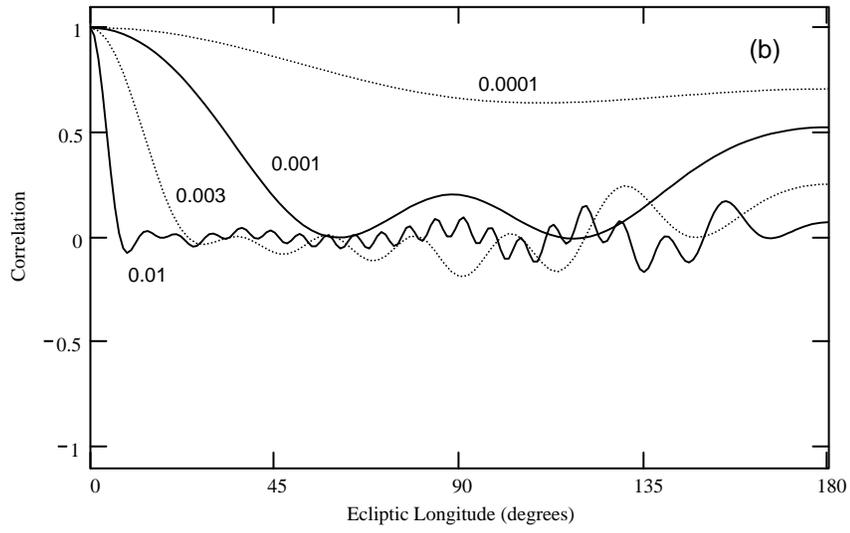
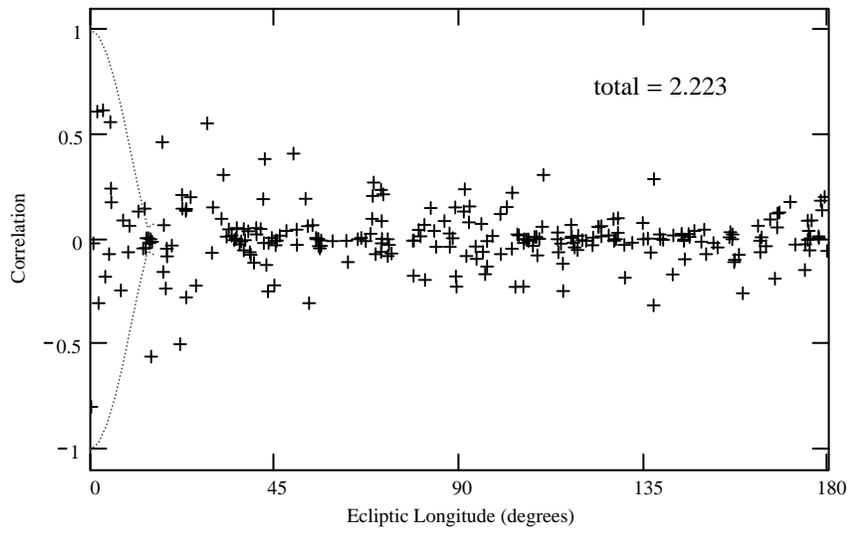
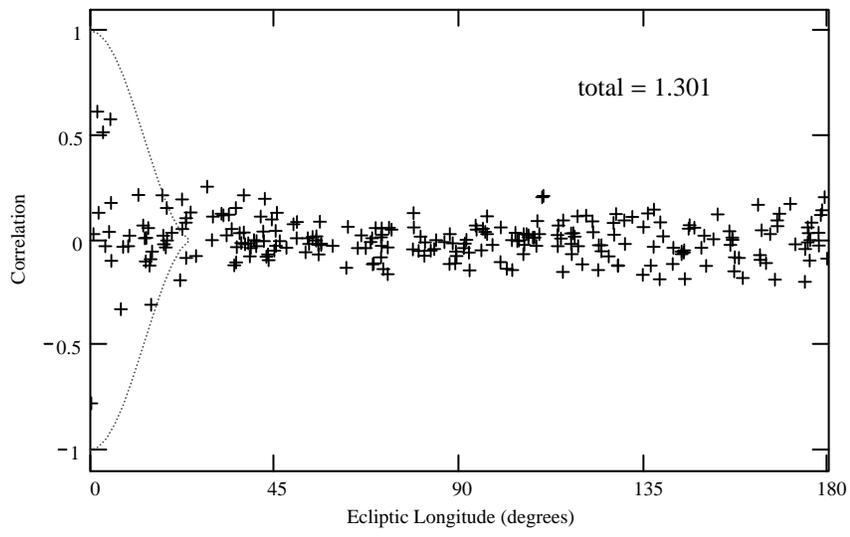
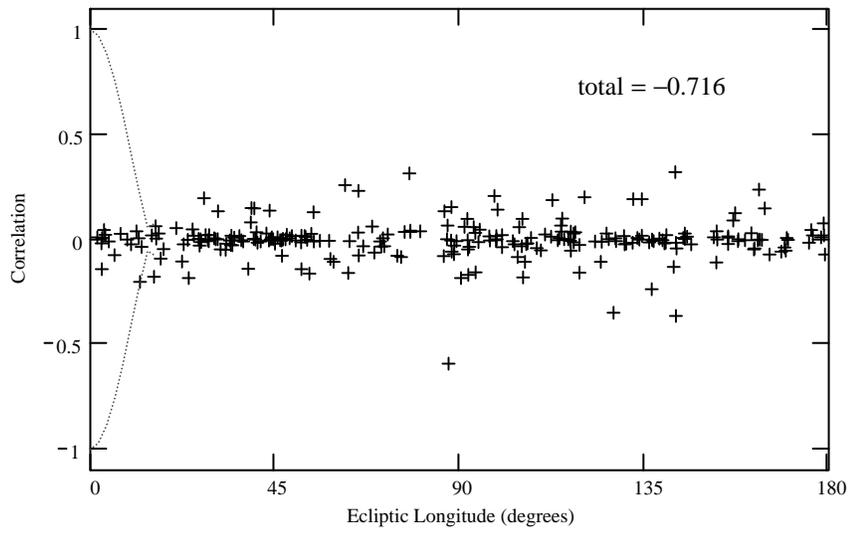
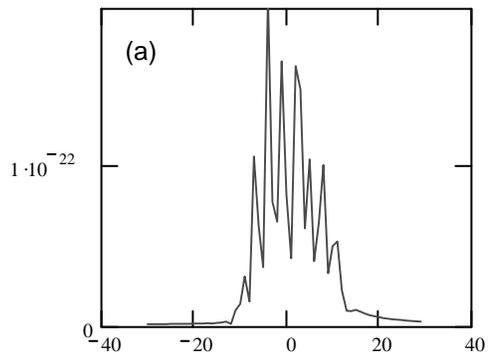
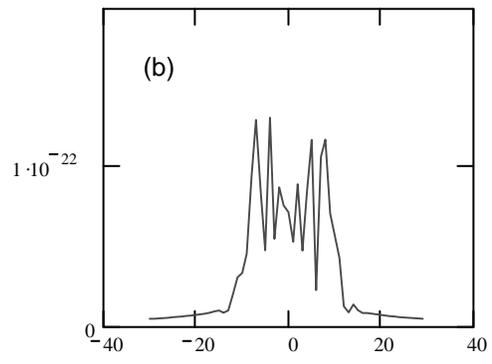
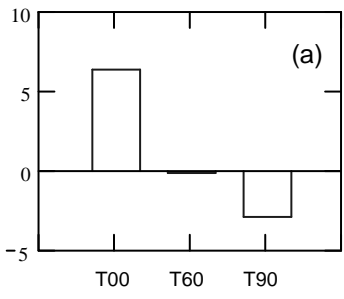
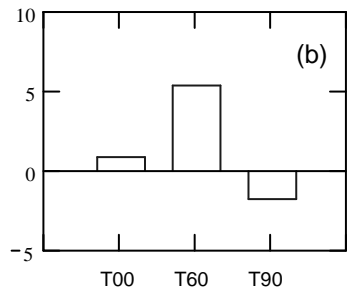
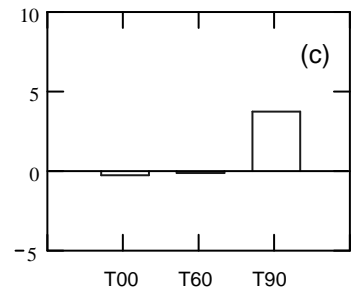
\bye